\documentclass[journal=jacsat,manuscript=article]{achemso}

\usepackage{chemformula} 
\usepackage[T1]{fontenc} 
\usepackage{comment}    

\author{Eleni Prountzou}
\email{eprountzou@ethz.ch}
\affiliation[Unknown University]
{ETH Zurich, Department of Physics, Institute for Quantum Electronics, Optical Nanomaterial Group, 8093 Zurich, Switzerland}
\author{Helena C. Weigand}
\author{Virginia Falcone}
\author{{\"U}lle-Linda Talts}
\author{Rachel Grange}

\title[An \textsf{achemso} demo]
  {Electro-Optic Modulation in Polycrystalline Barium Titanate Metasurfaces Enhanced by Poling
   }

\abbreviations{BTO, EO, SNIL, ITO, AC, DC, CMOS, FEM, FIB, PDMS, RIE}

\begin{document}

\begin{abstract}
    Electrically tunable metasurfaces leveraging the strong Pockel's effect in barium titanate (\ch{BaTiO3} or BTO) are a promising platform for reconfigurable free-space optical devices. However, the high cost, limited scalability, and restricted substrate compatibility of epitaxial BTO films hinder its exploitation. Here, we demonstrate free-space optical modulators based on imprinted BTO metasurfaces with targeted designs for optical and electric field confinement within the active material. With resonances exhibiting high quality factors of up to 200, we demonstrate improved transmission modulation at sub-volt driving amplitudes and frequencies up to 5 MHz. Additional enhancement is achieved via ferroelectric domain alignment, resulting in up to 25 \% higher modulation strength compared to the unbiased case and up to 75 \% compared to previous demonstrations. This enhanced EO response, arising from the effective permittivity engineering and domain orientation in these polycrystalline metasurfaces, holds significant potential for scalable and efficient EO modulators and active metasurfaces.
  
\end{abstract}

\section{Introduction}

Ferroelectric materials have recently attracted considerable interest due to their wide range of applications, including non-volatile memories \cite{Wang2025, LIU2025100870},
ferroelectric photovoltaics \cite{grinberg_perovskite_2013}, and acousto-optic or electro-optic (EO) devices \cite{doi:10.1021/acsphotonics.2c01589,
https://doi.org/10.1002/adma.202506790}. In telecommunications, where high-speed data transmission is essential, non-centrosymmetric materials with large EO coefficients and modulation bandwidths exceeding the kHz regime are required. While lithium niobate (\ch{LiNbO3}) is widely used in both integrated \cite{hu_integrated_2025} and free-space devices, reaching modulation speeds above GHz \cite{li_lithium_2020, doi:10.1021/acsphotonics.4c02354, di_francescantonio_efficient_2025}, barium titanate (\ch{BaTiO3} or BTO) has gained attention due to its striking EO coefficient of $r_{42}=1300$ pm/V (unclamped case) in bulk form \cite{https://doi.org/10.1002/adom.202001249}. However, the limited integrability and scalability of bulk BTO have driven research towards attaining BTO films via techniques such as molecular beam epitaxy, metal-organic chemical vapor deposition, radio frequency sputtering, pulsed laser deposition, and spalling \cite{reynaud_enhancement_2025, Tang:04, SZMYT2025102490, Sumon_2025, https://doi.org/10.1002/admi.202300665, doi:10.1021/acs.nanolett.5c02279}. Although epitaxial BTO on Si has achieved high-speed EO modulation with an EO coefficient of $r_{42}=923$ pm/V \cite{abel_large_2019}, epitaxial growth of layers remains energy-inefficient and restricted by substrate compatibility.

Conversely, wet-chemistry techniques, such as the sol-gel method, offer a versatile and scalable approach that is compatible with soft-nanoimprint lithography (SNIL). This etch-free bottom-up process enables the imprinting of nanostructures on a variety of substrates and has been used for the fabrication of linear and nonlinear metasurfaces \cite{doi:10.1021/cm071037m, https://doi.org/10.1002/smll.202304355, doi:10.1021/acs.nanolett.4c00711, https://doi.org/10.1002/adma.202418957}. Moreover, the polycrystallinity of these materials provides a simpler platform for designing and optically characterizing their effective polarization-independent EO performance \cite{10.1063/1.1819515}. However, the randomly oriented grains of BTO lead to a lower effective EO coefficient of approximately 27 pm/V \cite{https://doi.org/10.1111/jace.16815}, which, nevertheless, remains comparable to that of monocrystalline \ch{LiNbO3} \cite{https://doi.org/10.1002/pssa.200303911}.

Further increase of the EO response of polycrystalline materials can be achieved through permanently orienting their ferroelectric domains, a process known as poling. At ambient temperature, BTO retains its tetragonal phase exhibiting a spontaneous polarization along the c-axis (out-of-plane) due to the displacement of the \ch{Ti+} ions along this axis within each unit cell \cite{https://doi.org/10.1002/adom.202001249}. However, within a single crystal multiple ferroelectric domains exist and can posses a-axis (in-plane) or c-axis orientation. In BTO's tetragonal phase, polarization reversal occurs via the formation of opposing 180$^\circ$ or orthogonal 90$^\circ$ domains. The high EO coefficient of monocrystalline BTO has prompted the investigation of domain effects within the perovskite oxide with recent achievements in domain alignment \cite{Vasudevan:23, https://doi.org/10.1002/adom.202303058}, as well as periodic poling \cite{https://doi.org/10.1002/advs.202406248} of the crystal. Additionally, studies on ferroelectric switching in ultra-thin BTO films show that both epitaxial and polycrystalline BTO thin films can achieve coercive fields and remanent polarization values approaching those of the bulk material \cite{doi:10.1126/science.1103218, jiang_enabling_2022, https://doi.org/10.1002/aelm.202400440}. These advances are relevant for EO modulators, especially those based on polycrystalline BTO films and three-dimensional structures, since a stable remanent polarization can introduce a built-in birefringence that enhances the EO response without requiring continuous biasing.

In this context, polarization stability becomes an important figure of merit. While monocrystalline and epitaxial ferroelectrics such as \ch{BiFeO3} and \ch{PbTiO3} show retention periods of over a year \cite{zhang_superior_2020, doi:10.1021/acs.nanolett.5c04334, 10.1063/1.2794859}, retention in polycrystalline films is mainly governed by grain boundaries, local electric field inhomogeneity, and grain-size-dependent domain stability \cite{Song01072006, 10.1063/1.3499645, 10.1063/1.5081086}; factors that contrast with the polarization relaxation driven by domain wall defects observed in single-crystals. Solution-derived BTO allows for the tuning of these factors, as adjusting parameters such as the final annealing temperature can be used to optimize the grain size to prolong polarization retention \cite{falcone2026solutionderivedbariumtitanatewaveguides}. Despite the complex domain dynamics associated with polycrystallinity, BTO remains competitive for device integration due to its relatively high effective EO coefficient. To further enhance this Pockel's response, several studies in both single-crystalline and solution-processed polycrystalline BTO films, developed integrated EO modulators on lattice-matched buffer layers \cite{https://doi.org/10.1002/adma.202101128, https://doi.org/10.1002/adfm.202403024}. This growth provides directionality and minimizes stresses in the grown BTO film, rendering such platforms promising for future poling of EO devices.

Along the same direction of developing polycrystalline BTO-based modulators, a previous work reported a promising performance of an EO metasurface, showing potential for large-scale, and high-speed free-space devices \cite{doi:10.1021/acs.nanolett.4c00711}. The metasurfaces consisted of BTO pillars embedded in a \ch{SiO2} spacer layer. In this case, a glass substrate was used, with sandwiched top and bottom indium tin oxide (ITO) serving as transparent electrodes, enabling the characterization of the device's transmission modulation. By applying an alternating current (AC) voltage of 1.5 V at a driving frequency of 400 kHz, the measured amplitude modulation reached 0.04 \%. However, the electric field was stronger in the low permittivity \ch{SiO2} capping layer rather than the BTO pillars where the Pockel’s effect occurs. This suggests that optimized designs are required to further maximize the electric and optical field overlap within the nonlinear material.

Driven by the potential that polycrystalline BTO devices have showcased for nanophotonic devices, in this study, we focus on engineering the effective permittivity of a transparent free-space EO modulator based on solution-processed polycrystalline BTO metasurfaces fabricated via SNIL. By designing sandwich-type device architectures that support high electric fields inside the active material and operate at low driving voltages ($\leq$ 1.5 V), we enable ferroelectric domain alignment within nanostructured BTO, leading to enhanced transmission modulation. Furthermore, we show that EO modulation provides an indirect and accessible probe of the poling process and its temporal stability. These results establish a promising direction toward scalable and complementary metal-oxide-semiconductor (CMOS) compatible polycrystalline ferroelectric nanophotonic devices with enhanced EO performance.

\section{Results and discussion}

\begin{figure}[]
\centering\includegraphics[width=1\linewidth]{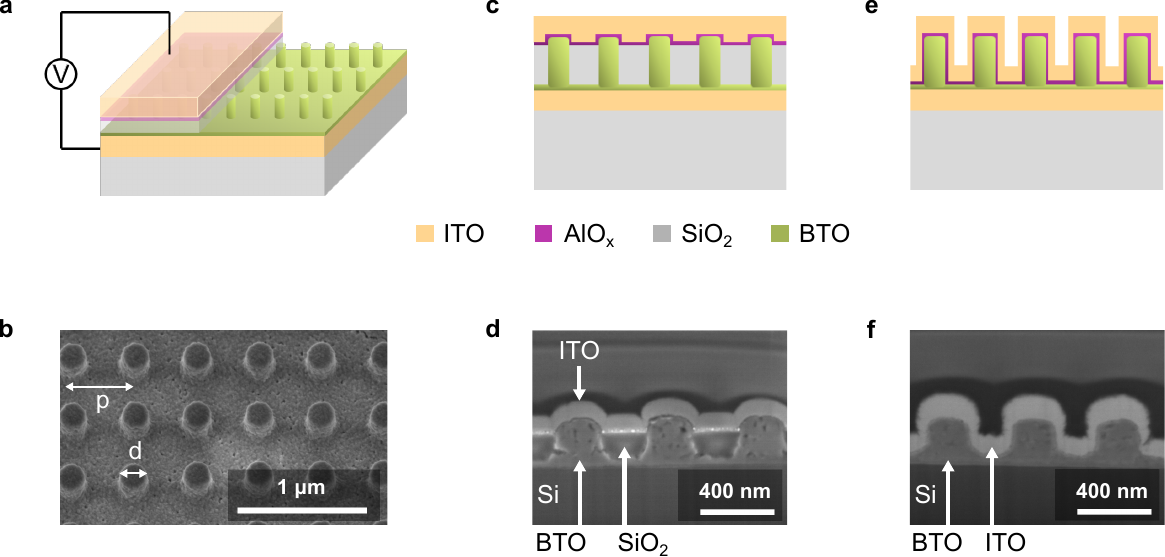}
\caption{Structural configurations of barium titanate (BTO) metasurfaces. (a) Schematic of the device used to demonstrate electro-optic modulation of the transmitted signal under an applied AC voltage across the sandwiched ITO electrodes. The BTO pillars are buried in a \ch{SiO2} layer and a conformal insulating $\mathrm{AlO_x}$ film is deposited on top of the capping \ch{SiO2} layer. (b) Scanning electron micrograph (SEM) (tilted at 30$^\circ$) showing the top view of the imprinted metasurface consisting of arrays of BTO pillars with a periodicity of p=500 nm and a diameter of d=200 nm. (c) Sketch of the embedded configuration, where \ch{SiO2} is etched to the height of the pillars. (d) Cross-sectional SEM of the embedded device. (e) Schematic cross-section of the conformal configuration, where $\mathrm{AlO_x}$ and ITO are deposited conformally over the nanostructures. (f) Cross-sectional SEM of the conformal device. All the cross-sectional SEM images were prepared using focused ion beam (FIB) milling. The devices shown in (d) and (f) are clone devices fabricated on silicon substrate without a ground ITO electrode below the metasurface for better imaging quality.}
\label{Figure 1}
\end{figure}

The transparent free-space EO modulators studied here feature sandwich-type electrode configurations (Figure \ref{Figure 1}(a)) that host resonant metasurfaces consisting of arrays of solution-processed BTO nanopillars (Figure \ref{Figure 1}(b)). The nanostructures are fabricated via the etchless SNIL process using an inverse polydimethylsiloxane (PDMS) mold on fused quartz substrates, chosen to ensure minimal transmission losses. The imprinted structures are subsequently annealed at $800^\circ C$ for 2 hours (fabrication and geometrical details in Experimental Section). More specifically, we investigate two distinct device architectures aiming to increase the electric field inside the BTO pillars. When the insulating planarization \ch{SiO2} layer is deposited in a thickness larger than the metasurface pillar height, the low permittivity \ch{SiO2} film captures most of the applied voltage, thereby diminishing the field in the high permittivity BTO \cite{doi:10.1021/acs.nanolett.4c00711}. To mitigate this effect, one approach is to reduce the \ch{SiO2} planarization layer to match the height of the pillars (Figures~\ref{Figure 1}(c), (d)), hereafter referred to as the embedded device. Another solution we propose is to entirely eliminate the \ch{SiO2} layer and instead conformally deposit the ITO electrode on top of the BTO nanostructures (Figures \ref{Figure 1}(e), (f)), hereafter referred to as the conformal device. In both approaches, a 20 nm-thick insulating layer of $\mathrm{AlO_x}$ is included between the BTO nanostructures and the top electrode to ensure electrical isolation during field application. Notably, both solutions maintain a low operating voltage of 1.5 V, as well as a small footprint of 50 x 50 $\mu m^2$, which renders them compatible with CMOS driver circuits \cite{soma_subvolt_2025}.

\begin{figure}
\centering\includegraphics[width=1\linewidth]{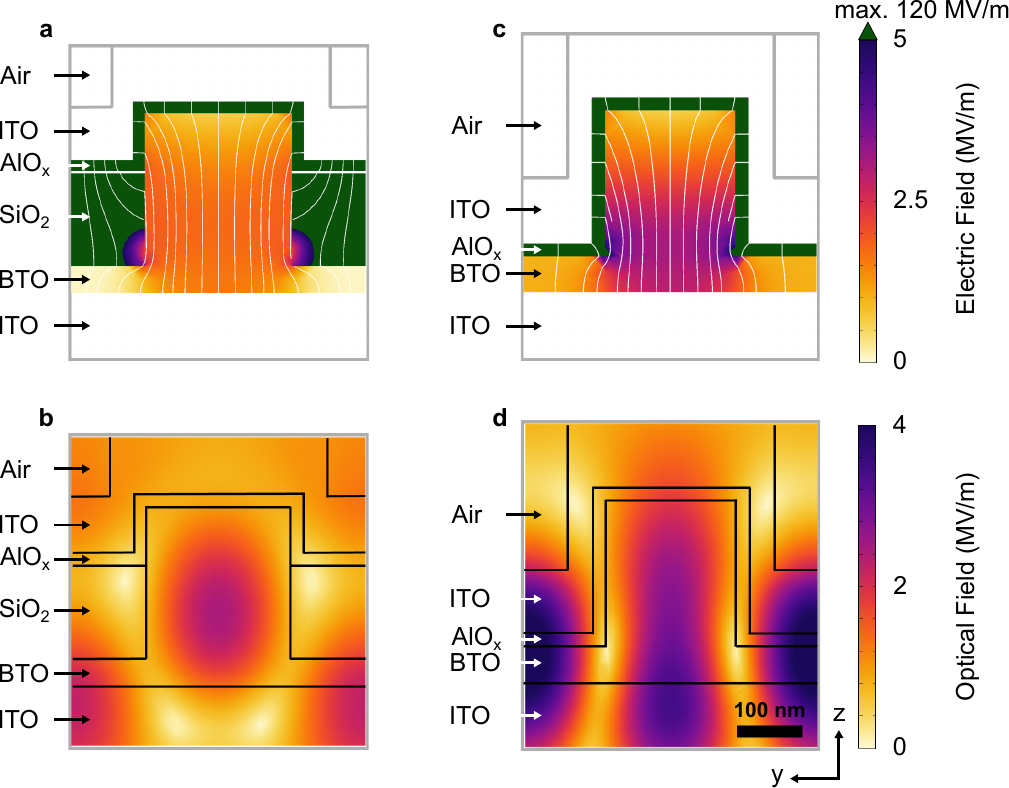}
\caption{Electric and optical field simulations for the two electro-optic device configurations. Panels (a) and (b) show finite-element simulations of the electric and optical fields, respectively, for the embedded structure at 783.6 nm. Panels (c) and (d) correspond to the electric and optical field distributions, respectively, of the conformal structure at 768.6 nm. The electric field lines are overlaid in white in panels (a) and (c). The color scales for the electric field ((a), (c)) and the optical field ((b), (d)) are shown on the right for both device configurations. For the electric field, the color scale is saturated at 5 MV/m, with a maximum field strength of 120 MV/m in $\mathrm{AlO_x}$ and \ch{SiO2} (in green). The scale bar and coordinate axes shown at the bottom right of panel (d) are global and common for all simulations.}
\label{Figure 2}
\end{figure}

To validate the effect of the device architecture on the EO performance, finite element method (FEM) simulations were performed for the two devices. The radius and periodicity of the meta-atoms were adjusted to position their resonances within the 750-790 nm range, corresponding to the tunable laser used for the transmission modulation measurements. Due to differences in the device stacks, separate optimizations were required for each configuration (simulation details, including relative permittivity values, in SI section 1). The electric and optical field distributions for the embedded and conformal devices at the optimal wavelengths of 783.6 and 768.6 nm, respectively, are shown in Figures~\ref{Figure 2}(a)-(d). A positive potential of 1.5 V is applied on the top electrode for both devices to investigate the electric field distribution and estimate the refractive index change ($\Delta n$) in BTO. However, it is noted that from the applied 1.5 V, only 0.06 V actually drop across the active material in the two devices, while the greater part of it falls on the $\mathrm{AlO_x}$ and \ch{SiO2} layers (Figures~\ref{Figure 2}(a), \ref{Figure 2}(c)). This further confirms that minimizing the proportion of these insulating layers is essential in order to increase the applied field in the metasurfaces.

An optimal EO performance is supported by a spatial overlap that satisfies the electric and optical field distributions, which can be quantified by the overlap integral factor \cite{doi:10.1021/nl404513p}. The combination of the electric and optically induced fields exhibits a mode overlap integral of 27 \% for the embedded device and 37 \% for the conformal configuration. In addition, a strong electric field intensity in the BTO pillar indicates a corresponding increase in $\Delta n$, considering the proportionality between $\Delta n$ and the applied electric field $E$, as given by the relation: $\Delta n=-\frac{1}{2} \cdot r_{eff} \cdot {n^3} \cdot E$, where $r_{eff}$ is the effective EO coefficient and $n$ the effective refractive index of the material. In the case of the embedded nanostructures, the simulation (Figure~\ref{Figure 2}(a)) predicts a 10-fold enhancement in the electric field intensity in the BTO pillar compared to previous device \cite{doi:10.1021/acs.nanolett.4c00711}, while for the conformal structure, the simulation shows a 13-fold enhancement (Figure~\ref{Figure 2}(c)). Therefore, the simulated enhanced fields are expected to yield a comparable increase in $\Delta n$ and, consequently, in the modulation efficiency of the devices. Considering an average refractive index of 1.91 for our polycrystalline BTO in the wavelength range of 750-790 nm (as determined through ellipsometry measurements available in previous study 
\cite{doi:10.1021/acs.nanolett.4c00711}), and an effective EO coefficient of 27 pm/V \cite{https://doi.org/10.1111/jace.16815}, the expected refractive index shift at the center of the BTO pillars is calculated to be approximately $\Delta n_{embed} = 1.8 \times 10^{-4}$ for the embedded configuration ($E_{embed,~BTO}=1.9~MV/m$) and $\Delta n_{conf} = 2.3 \times 10^{-4}$ for the conformal structure ($E_{conf,~BTO}=2.4~MV/m$).

\begin{figure}
\centering\includegraphics[width=1\linewidth]{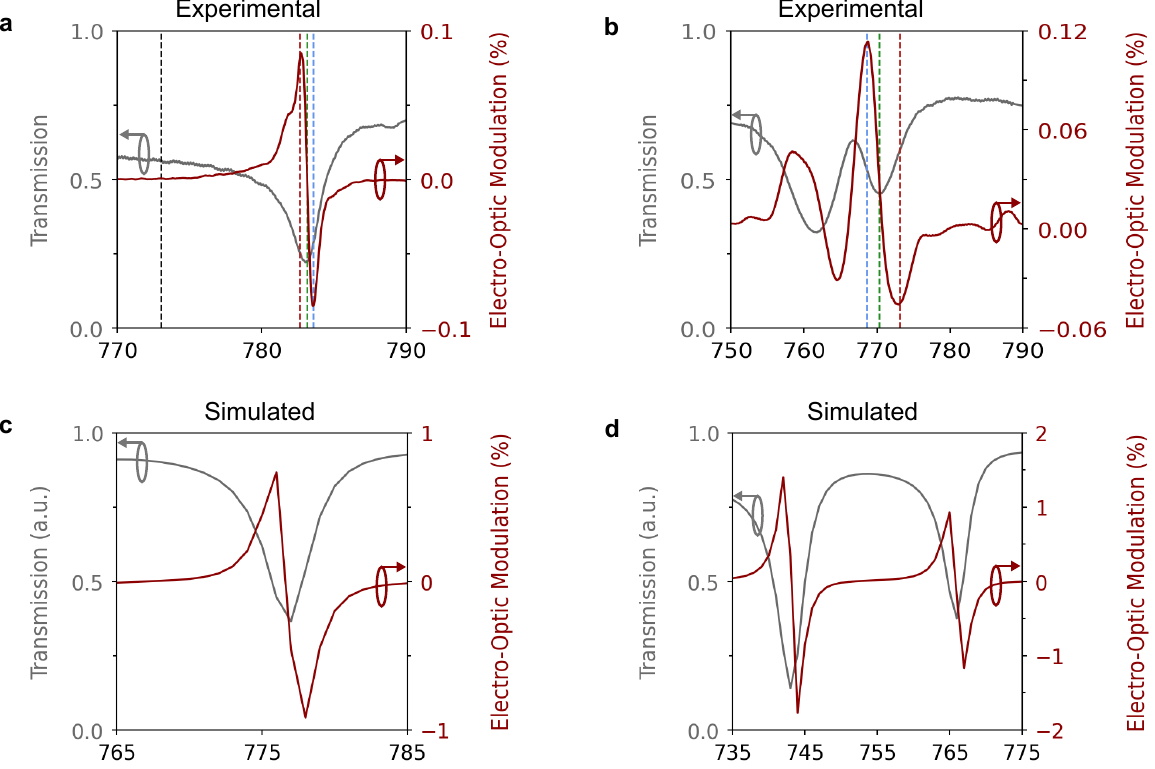}
\caption{Experimental and simulated results for the embedded and conformal device configurations. Panels (a) and (b) illustrate the measured transmission spectrum (in gray) and electro-optic (EO) modulation (in red) of the embedded and conformal structures, respectively, when an electric field with an amplitude of 1.5 V and a driving frequency of 400 kHz are applied. The transmission spectra were obtained by sweeping the laser wavelength and acquire the DC component of the demodulated signal ($T_{V_{DC}=0}$). The transmission data are normalized by the transmission of an unstructured region. The EO modulation is defined as the ratio $\frac{\Delta T}{T} = \frac{T_{V=V_{pp}} - T_{V=0}}{T_{V=0}}$, where $V_{pp}$ is the peak-to-peak amplitude of the sinusoidal wavefunction applied at a specific frequency. It is noted that the negative values reported for the EO efficiency stem from a sign change in the transmission signal relative to the unbiased signal. Therefore, only the absolute value of the EO modulation should be considered. The vertical dotted lines in the plots mark the different wavelengths at which the EO response was further examined in Figures S4(a)-(d) in the SI section 4. Panels (c) and (d) present the finite-element simulated transmission spectra (in gray) and EO modulations (in red) of the embedded and conformal architectures, respectively.}
\label{Figure 3}
\end{figure}

To confirm the promising simulated results, the two metasurface-based free-space modulators were experimentally characterized by measuring the collinear transmission spectra and transmission amplitude modulation in normal incidence. A linearly polarized continuous-wave laser tunable in the range of 750-790 nm was utilized to probe the devices and the signal was collected using a 50x objective and directed to both a camera (for imaging the sample) and a photodiode (for signal detection) via a 50/50 beam splitter. The photodiode output signal was guided to a lock-in amplifier, which simultaneously provided the driving AC frequency and amplitude for the electric field applied to the electrodes, while demodulating the transmitted signal (detailed in ~\cite{doi:10.1021/acs.nanolett.4c00711}).

Starting with the embedded configuration, a metasurface composed of 125-nm-radius cylindrical pillars arranged in a square lattice with 500 nm period was investigated. The experimentally measured resonance, shown in Figure~\ref{Figure 3}(a) (gray line), exhibits a quality (Q) factor of 200 and a linewidth of 4 nm, as determined by fitting a Fano profile to the experimental data, which is in good agreement with the value of 207 calculated for the simulated resonance (fit details in SI section 2). The Fano model was selected due to the asymmetric lineshape, attributed to coupling between lattice and Mie modes forming hybrid resonances \cite{doi:10.1021/acs.nanolett.4c00711}. In both configurations, the data presented in Figures \ref{Figure 3}(a) and \ref{Figure 3}(b) were processed to suppress the Fabry-Pérot oscillations within the fused quartz substrate, which otherwise reduce spectral clarity (details for the process in SI section 3). In the same plot (Figure~\ref{Figure 3}(a)), the EO modulation (red line) is shown as a function of the wavelength. The evaluation of the EO modulation efficiency is determined through the calculation of the relative transmission amplitude modulation as the ratio $\frac{\Delta T}{T} = \frac{T_{V=V_{pp}} - T_{V=0}}{T_{V=0}}$ when a sinusoidal voltage waveform with a peak-to-peak amplitude of $V_{pp}$ is applied at a given frequency. The EO efficiency reaches its maximum absolute value of 0.08 \% at the steepest transmission resonance slope, that is, when $ dT/d\lambda $ is largest.

For the conformal configuration to have a resonance in the near infrared spectral range, the metasurface consisted of meta-atoms with 450 nm periodicity and 100 nm radius. Figure \ref{Figure 3}(b) shows the experimental EO modulation (red line) and the hybrid resonance (gray line), with a Q-factor of approximately 115 and a linewidth of 6.5 nm, which is in the same order of magnitude with the value of 204 calculated from the simulations (fit details in SI section 2). In this design, the absence of the \ch{SiO2} layer increases scattering losses from the surface granularity of BTO, leading to a lower Q-factor. In contrast, embedding in glass strengthens the lattice resonant effects through phase-matched in-plane diffraction in the substrate and surrounding media and mitigates these roughness-induced losses \cite{10.1063/1.5094122}. Nevertheless, the conformal device still achieves a stronger modulation of 0.12 \%, as the elimination of the low permittivity \ch{SiO2} layer enables a higher electric field inside the BTO nanopillars.

The experimental EO modulation values are compared against FEM simulations, which predict transmission modulations of 0.9 \% and 1.8 \% for the embedded (Figure \ref{Figure 3}(c)) and conformal (Figure \ref{Figure 3}(d)) designs, respectively. The deviations from the experimental results likely arise from fabrication-related imperfections (e.g., multiple scattering from surface roughness, porosity, high conductivity). Despite the 20 nm-thick $\mathrm{AlO_x}$ coating, the insulation remains inadequate due to the inherent porosity of the solution-derived metal oxide, which is a result of the evaporation of organic sub-products during annealing. The resistance of the devices was measured to be approximately in the order of 10 k$\Omega$, far from the expected M$\Omega$ range that would ensure sufficient insulation. To reduce the conductivity of the devices and increase the effective electric field in the metasurfaces, the porosity of the BTO sol-gel must be targeted, as increasing the thickness of either of the low permittivity insulating layers (\ch{SiO2} or $\mathrm{AlO_x}$) would instead reduce the field strength inside the BTO pillars (Figures \ref{Figure 2}(a), (c)) \cite{doi:10.1021/acs.nanolett.4c00711}. Alternatively, replacing \ch{SiO2} with a higher relative permittivity oxide, such as \ch{TiO2}, could reduce the electric field proportion in the insulator. However, it is imperative to emphasize that these results represent a substantial advancement compared to prior reported values for EO BTO metasurface modulators, with the embedded configuration achieving twice and the conformal design three times higher EO modulation \cite{doi:10.1021/acs.nanolett.4c00711}.

\begin{figure}
\centering\includegraphics[width=1\linewidth]{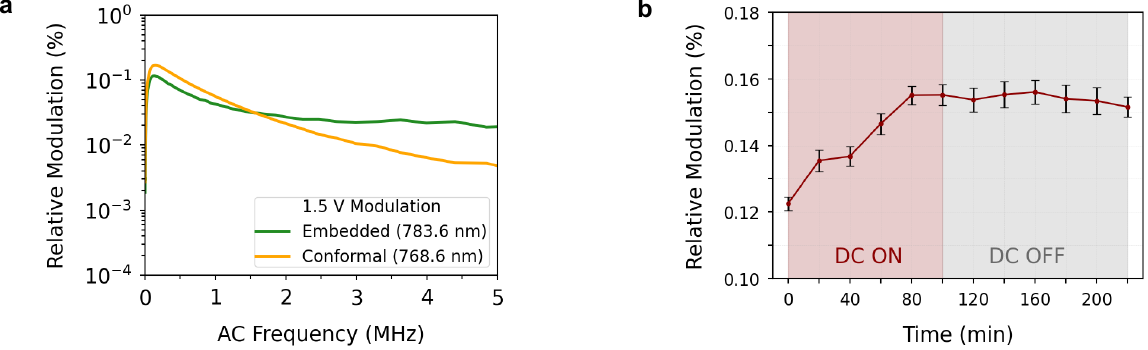}
\caption{Experimental electro-optic performance of the embedded and conformal devices. (a) Relative modulation of both devices as a function of the applied AC driving frequency for a fixed amplitude of 1.5 V, measured at the optimal operating wavelengths of each device. (b) Temporal evolution of the relative modulation of the conformal device under a DC bias of an absolute value of 10 V (33 MV/m) applied for 100 minutes, during which ferroelectric domains alignment enhances the electro-optic modulation. After the bias is removed, the modulation remains stable during a two hours relaxation period.
}
\label{Figure 4}
\end{figure}

To further assess the devices' performance, the dependence of the relative modulation on both voltage amplitude (at 400 kHz) (Figures S4(a), S4(b) in SI section 4) and frequency (at 1.5 V) (Figure \ref{Figure 4}(a)) was examined for the two devices at the wavelengths corresponding to their maximum transmission modulation (additional data for all examined wavelengths are provided in SI section 4). In both cases, within the low voltage regime ($\leq$ 1.5 V), the modulation scales linearly with the AC amplitude (Figures S4(a), S4(b) in SI section 4), confirming the Pockel's effect as the primary modulation mechanism. In addition, our devices yield up to 600 times greater modulation efficiency than their equivalent unpatterned films (data available in SI section 4). Complementary to the modulation amplitude response, our free-space EO modulators exhibit frequency bandwidths exceeding the kHz range, highlighting their potential for high speed EO modulation (Figure \ref{Figure 4}(a)). It is evident that a strong modulation occurs at low frequencies, with a 3 dB drop near 1 MHz for both devices; however, for the embedded configuration, it remains stable up to 5 MHz. This frequency limitation originates not from the intrinsic EO effect, but from the RC limitations of the device stack (electric circuit characterization in SI section 5). Improving the electrodes conductivity or reducing the capacitor area could therefore extend the modulation bandwidth \cite{doi:10.1021/acs.nanolett.4c00711, doi:10.1021/acsphotonics.1c01582}.

Beyond improving the device design for increased effective electric fields, permanent poling in polycrystalline ferroelectrics would allow a lasting modification of their birefringence, increasing the effective EO coefficient without the need for continuous DC biasing. The device architectures demonstrated here enable the alignment of ferroelectric domains in the nanostructured polycrystalline BTO, leading to enhanced EO performance and demonstrating sustained stability. Figure \ref{Figure 4}(d) illustrates the evolution of the relative modulation when a DC bias of an absolute value of 10 V (33 MV/m) is applied to the conformal device for 100 minutes, with the response recorded in intervals of 20 minutes by instantaneously turning off the DC voltage and record the device's AC response with a driving frequency of 400 kHz and an amplitude of 1.5 V (details for the embedded structure in SI section 6). While a static electric field can drive the motion of both 90$^\circ$ and 180$^\circ$ domain walls, permanent polarization is generally associated with the switching of 180$^\circ$ domains, which requires sufficiently long poling durations to enable their motion \cite{https://doi.org/10.1111/j.1151-2916.1956.tb15623.x, 10.1063/1.1722606}. The reconfiguration of domains was observed to be saturated after 100 minutes of applying static electric field achieving 25 \% increase in modulation amplitude. This enhancement arises from the change in the birefringence of BTO, which results in a higher effective EO coefficient and therefore improves the relative modulation. As no additional enhancement in modulation amplitude was observed after 100 minutes, the DC bias was removed, and the domains were let to relax. Following the poling process, the temporal stability of the device was monitored for 120 minutes. During this period, the device maintained stable modulation for about one hour, after which a slight decline was observed. Such stability is challenging to achieve in this material due to the complex dynamics of its ferroelectric domains, which can backswitch upon removal of the bias, especially when the applied field is not sufficient to fully switch them \cite{10.1063/1.3212975}. These results indicate that poling three-dimensional polycrystalline solution-derived BTO nanostructures is feasible and leads to an improved EO performance with prolonged stability.

While long polarization retention has been extensively studied in monocrystalline epitaxial ferroelectric thin films \cite{zhang_superior_2020, doi:10.1021/acs.nanolett.5c04334, 10.1063/1.2794859}, these investigations have mainly focused on planar geometries. In contrast, the direct investigation of poling and domain stability in nanostructured and metsurface-based polycrystalline materials remains experimentally challenging. In this work, EO modulation is employed as an indirect probe of ferroelectric domain reconfiguration in BTO polycrystalline nanostructures, enabling the assessment of polarization stability in metasurfaces. Further increase could be achieved by improving the device resistance and electrode conductivity, which will allow the application of higher DC voltages and thereby promote stronger domain alignment. In addition, another promising direction involves poling the devices at elevated temperatures above the Curie point \cite{Roseman01071998, https://doi.org/10.1002/pssa.201532373, Shibiru_2024}. Moreover, in monocrystalline epitaxial ferroelectric films, long polarization retention has been achieved through domain wall pinning strategies, such as the introduction of nanoscale defects that apply a local compressive strain throughout the film stabilizing the switched polarization \cite{zhang_superior_2020}. For epitaxial BTO, retention has been shown to depend on film thickness, with thinner layers ($\approx$ 10 nm) exhibiting longer stability \cite{10.1063/1.5020549}, while growing BTO on lattice-matched buffer layers were shown to improve the EO performance of integrated modulators \cite{https://doi.org/10.1002/adma.202101128, https://doi.org/10.1002/adfm.202403024}. In this context, doping of BTO and the use of transparent lattice-matched substrates, such as \ch{MgO}, or suitable buffer layers represent promising routes to improve stability in EO devices.

\subsection{Conclusions}

Resonant nanostructured BTO metasurfaces fabricated via the SNIL method have been shown to significantly enhance EO modulation compared to plain thin films, offering a scalable and etch-free route for optical devices \cite{doi:10.1021/acs.nanolett.4c00711}. Although their modulation speed and electric field strength are superior to those of liquid crystals \cite{doi:10.1021/acsnano.3c04071} and phase change materials \cite{https://doi.org/10.1002/advs.202413316}, their modulation amplitude was partly limited due to the electric field being stronger in the \ch{SiO2} planarization layer than in the BTO pillars.

In this work, to overcome this limitation we develop two improved device configurations: an embedded design, where the \ch{SiO2} layer is etched up to the height of the pillars, and a conformal design, where the \ch{SiO2} capping layer is entirely omitted. These two configurations not only enable up to three times higher EO modulation compared to previous work \cite{doi:10.1021/acs.nanolett.4c00711}, but also allow the poling of the BTO domains, resulting in a maximum increase of 25 \% of the relative modulation compared to the unbiased state, and a total improvement of 75 \% over previously reported value \cite{doi:10.1021/acs.nanolett.4c00711}. Moreover, the conformal device exhibits polarization stability for up to two hours, indicating the potential for sustained EO enhancement without the need for continuous DC biasing. Long-term retention reported in polycrystalline PZT, exceeding one year, further reinforces the potential of polycrystalline BTO for EO applications \cite{10.1063/1.3499645}. Finally, by applying a driving voltage of only 1.5 V, we enable the energy-efficient integration of these devices with CMOS-compatible technologies.

Further optimization of the efficiency could be achieved via the implementation of high-Q-factor resonances, such as quasi-bound states in the continuum, to maximize extinction ratios \cite{Shen:24, TIAN2025131181}. For polycrystalline BTO, reducing the porosity would not only raise the material's effective refractive index, thereby relaxing current design constraints for high-Q resonances, but would also improve the insulating properties of the nanostructures, enabling electric-field distributions closer to those predicted by simulations. In this context, the SNIL method applied here offers a highly adaptable fabrication platform that is well suited for a broad range of flat optical components. Recent demonstrations of metalenses based on solution-derived ferroelectrics \cite{Weigand:22, https://doi.org/10.1002/adma.202418957} highlight the potential of this approach, while the enhanced modulation efficiency achieved in this work provides a practical route toward developing high-speed, scalable EO metasurfaces and BTO-based metalenses, which has primarily been the subject of theoretical studies to date \cite{
nano11082023, Lv:22, Wang:25}.


\section{Experimental}

The complete process for the synthesis of the solution-derived BTO and its imprinting by SNIL, as well as the optical setup for the electro-optic characterization of the devices, can be found in previous work \cite{doi:10.1021/acs.nanolett.4c00711}.
Regarding the fabrication process, the cross-section of both devices is illustrated in Figures~\ref{Figure 1}(c), (e). The 100 nm-thick ITO electrodes were sputtered in high vacuum on top of the 1 mm- and 500 $\mathrm{\mu m}$-thick fused quartz substrate for the embedded and conformal devices, respectively. Subsequently, the BTO sol-gel was spin-coated for 5 seconds at 1000 rpm and the metasurfaces were imprinted using a soft polydimethylsiloxane (PDMS) mold. After a 2-hour curing of the samples at $70^\circ C$, the molds were removed and the samples were annealed at $800^\circ C$ for 2 hours with a ramp rate of $2^\circ C/min$ to allow the BTO to transit from the amorphous to the polycrystalline phase. In the embedded configuration (Figure~\ref{Figure 1}(c)), a 500 nm-thick spin-on glass was added as a capping layer to the BTO pillars and was heat-treated at $400^\circ C$ for 30 minutes. After that, reactive ion etching (RIE) of approximately 290 nm of excess \ch{SiO2} on top of the BTO meta-atoms was followed, until the protruding pillars appeared, as shown in Figure~\ref{Figure 1}(d). The RIE was performed at 20 mTorr using a \ch{CHF3/Ar} plasma gas. Finally, on both devices, a 20-nm $\mathrm{AlO_x}$ layer was deposited by atomic layer deposition, followed by the 100 nm-thick top electrode that was conformally sputtered on selected areas as the final step in the devices fabrication. The selective deposition of the ITO top electrodes was achieved using photolithography to pattern the desired areas and a lift-off process.

The metasurfaces are composed of cylindrical pillar-shaped unit cells with engineered resonances across the visible and near-infrared spectra, which can be achieved by sweeping the periodicity (400 - 600 nm) and pillar diameters (100 - 300 nm). The pillars have a height of approximately 260 and 225 nm and a residual layer of 45 and 55 nm for the embedded and conformal devices, respectively. As previously reported \cite{doi:10.1021/acs.nanolett.4c00711}, the volumetric shrinkage of the imprinted nanostructures imposes a 0.5 filling factor limit to counteract this effect. The aspect ratios achieved here are close to 1.

\begin{acknowledgement}

The authors thank the Scientific Center for Optical and Electron Microscopy (ScopeM), the Binning and Rohrer Nanotechnology Center (BRNC), and the FIRST cleanrooms of Eidgenössische Technische Hochschule (ETH) Zurich, as well as Agostino Di Francescantonio for technical and methodological assistance, Alfonso Nardi for his contributions and discussions regarding the electrical circuit calculations, and Prof. Dr. Morgan Trassin for the insightful discussions on the ferroelectric domain alignment of our material. This work was supported by the Swiss National Science Foundation SNSF (Consolidator Grant 213713).

\end{acknowledgement}

\begin{suppinfo}

The Supporting Information includes:
\begin{itemize}
  \item[] Details of the simulations, Fano fitting of the resonances, Fabry-Pérot oscillations within the substrate cavity, AC response of the devices, electrical circuit characterization, and electro-optic response of the embedded device under DC biasing.
\end{itemize}

\end{suppinfo}

\bibliography{references_arxiv}

\end{document}